\documentclass[12pt]{article}

\usepackage{graphicx}
\usepackage{dcolumn}
\usepackage{bm}
\usepackage{mathbbol}

\begin{document}
\begin{center}
\textbf{Hawking radiation of black holes in the $z = 4$
Horava-Lifshitz gravity}
\end{center}
\begin{center}
 De-You Chen \footnote{
Email: deyouchen@uestc.edu.cn}, Haitang Yang \footnote{ Email:
hyanga@uestc.edu.cn} and Xiao-Tao Zu
\\School of Physical Electronics, University of
Electronic Science and Technology of China, Chengdu, 610054, China.\\
\end{center}

\textbf{Abstract:} We investigate the Hawking radiation of (3+1)-
and (4+1)-dimensional black holes in the $z = 4$ Horava-Lifshitz
gravity with fermion tunnelling. It turns out that the Hawking
temperatures are recovered and are in consistence with those
obtained by calculating surface gravity of the black holes. For the
(3+1)-dimensional black holes, the Hawking temperatures are related
to the fundamental parameters of Horava-Lifshitz gravity.



\textbf{1. Introduction}

Motivated by the Lifshitz theory, a different field theory model for
a UV complete theory of gravity was recently proposed by Horava
[1-4]. This model is based on a scaling symmetry which treats space
and time differently. It has the same dynamical degrees of freedom
as in general relativity, but without the full diffeomorphism
invariance. The power counting renormalizability of this theory
indicates that fundamental theory could be specified by a finite
number of UV parameters which are in principle measurable in the IR.
In this theory, the divergence of the speed of light in the UV may
resolve the horizon and flatness problems. Moreover, this model is
helpful to solve some singularity problems, for instance, black hole
singularity and cosmological singularity. Since it was proposed,
this Horava-Lifshitz model has attracted considerable attention. In
the literature, the properties of Horava-Lifshitz gravity have been
extensively studied. The black hole solutions in the $z = 3$ and $z
= 4$ cases were obtained in [5-10] and [11], respectively. The
cosmology solutions of the theory were addressed in [12-17],
cosmological perturbation and the related properties in the theory
was discussed in [18-21]. The properties of black hole solutions
were researched in [22-26] and other properties of Horava-Lifshitz
gravity were investigated in [27-39].

The research on the $z = 3$ Horava-Lifshitz gravity has been carried
out thoroughly. In [11] and the references therein, it is believed
that the new massive gravity can be used to construct a $z = 4$
Horava-Lifshitz gravity and the spectral dimension calculated using
the numerical Causal Dynamical Triangulations approach for quantum
gravity in (3+1)-dimensions prefers the z = 4 Horava-Lifshitz
gravity in the UV. With this reason, we investigate the
thermodynamic properties of (3+1)-dimensional and 4+1 dimensional
black holes in the $z = 4$ Horava-Lifshitz gravity by fermion
tunnelling in this Letter. Recently the Hawking radiation of general
black holes has been studied [40-63]. In Refs.[40, 41], the Hawking
radiation of has been researched by Parikh-Wilczek's tunnelling
method and anomaly cancellation method, respectively. In Ref.[43],
the Hawking radiation has been investigated by Hamilton-Jacobi
method. The black hole radiates not only scalar particles but also
Dirac particles. Subsequently the tunnelling effect of spin 1/2
particles in the static spacetime was first studied in Ref.[51] by
Hamilton-Jacobi method, and this work was extended to
four-dimensional spacetime and higher-dimensional spacetime [52-63].

This Letter is organized as follows. In sect. 2, the solution of
(3+1)-dimensional black holes in the $z = 4$ Horava-Lifshitz gravity
is reviewed and the Hawking radiation of the black holes is
discussed with fermion tunnelling. In sect. 3, the Hawking radiation
of fermions in (4+1)-dimensional black holes in the $z = 4$
Horava-Lifshitz gravity is studied and the Hawking temperatures are
reproduced. Sect. 4 is devoted to discussions and conclusion.

\textbf{2. Hawking radiation of (3+1)-dimensional black holes in the
$z = 4$ Horava-Lifshitz gravity}

In this section, we investigate the Hawking radiation of
(3+1)-dimensional black holes in the $z = 4$ Horava-Lifshitz gravity
theory. We first review the solution of black holes that was
obtained  by Cai et. al. [11]. According to the ADM decomposition of
the metric, the (D+1)-dimensional metric can be written as

\begin{equation}
\label{eq1} ds^2 = - N^2c^2dt^2 + g_{ij} \left( {dx^i - N^idt}
\right)\left( {dx^j - N^jdt} \right),
\end{equation}

\noindent where $i = 1, \cdots ,D$ and $c$ is the speed of light.
The kinetic term of the action is given by

\begin{equation}
\label{eq2} S_K = \frac{2}{\kappa ^2}\int {dtd^Dx\sqrt g N\left( {K{
}_{ij}K^{ij} - \lambda K^2} \right)} ,
\end{equation}

\noindent where $K_{ij} = \frac{1}{2N}\left( {\partial _t g_{ij} -
\nabla _i N_j - \nabla _j N_i } \right)$, $\kappa $ is the coupling
constant and $\lambda$ is a dimensionless parameter and should be
equal to 1 in the IR to restore general relativity here. From the
detailed balance principle, the potential term is obtained as

\begin{equation}
\label{eq3} S_V = \frac{\kappa ^2}{8}\int {dtd^Dx\sqrt g
NE^{ij}\varsigma _{ijkl} E^{kl}} ,
\end{equation}

\noindent with $E^{ij}$ coming from a D-dimensional relativistic
action in the form $E^{ij} = \frac{1}{\sqrt g }\frac{\delta W_D
\left[ {g_{ij} } \right]}{\delta g_{ij} }$, and $\varsigma _{ijkl} =
\frac{1}{2}\left( {g_{ik} g_{jl} + g_{il} g_{jk} } \right) - \tilde
{\lambda }g_{ik} g_{jl} $, $\tilde {\lambda } = \frac{\lambda
}{D\lambda - 1}$. Using the kinetic term and potential term, the
action of the (3+1)-dimensional Horava-Lifshitz theory is expressed
as

\[\mathcal {L}= \mathcal {L}_0 + \mathcal {L}_1 ,\]

\[
\mathcal {L}_0 = \sqrt g N\left\{ {\frac{2}{\kappa ^2}\left(
{K^{ij}K{ }_{ij} - \lambda K^2} \right) + \frac{\kappa ^2\mu
^2\left( {\Lambda _W R - 3\Lambda _W^2 } \right)}{8\left( {1 -
3\lambda } \right)}} \right\},
\]

\[
\mathcal {L}_1 = - \sqrt g N\frac{\kappa ^2}{8}\left\{
{\frac{4}{\omega ^4}C^{ij}C_{ij} - \frac{4\mu }{\omega
^2}C^{ij}R_{ij} - \frac{4}{\omega ^2M}C^{ij}L_{ij} + \mu ^2G_{ij}
G^{ij} + \frac{2\mu }{M}G^{ij}L_{ij} } \right.
\]

\begin{equation}
\label{eq4} \left. { + \frac{2\mu }{M}\Lambda _W L +
\frac{1}{M^2}L_{ij} L^{ij} - \tilde {\lambda }\left(
{\frac{L^2}{M^2} - \frac{\mu L}{M}\left( {R - 6\Lambda _W } \right)
+ \frac{\mu ^2R^2}{4}} \right)} \right\} ,
\end{equation}

\noindent where
\[
C^{ij} = \varepsilon ^{ikl}\nabla _k \left( {R_l^j -
\frac{1}{4}R\delta _l^j } \right), G^{ij} = R^{ij} -
\frac{1}{2}g^{ij}R,
\]

\[
L^{ij} = \left( {1 + 2\beta } \right)\left( {g^{ij}\nabla ^2 -
\nabla ^i\nabla ^j} \right)R + \nabla ^2G^{ij}+\]

\[+ 2\beta R\left( {R^{ij} - \frac{1}{4}g^{ij}R} \right) + 2\left(
{R^{imjn} - \frac{1}{4}g^{ij}R^{mn}} \right)R_{mn},
\]

\[L \equiv g^{ij}L_{ij} = \left( {\frac{3}{2} + 4\beta }
\right)\nabla ^2R + \frac{1}{2}\beta R^2 + \frac{1}{2}R^{ij}R_{ij}
,\] $\mu $, $\omega ^2$, $M$ and $\beta $ are coupling constants. To
get the explicit black hole solution, the value of $\beta $ is
chosen as -3/8 in this section [11]. In order to obtain the general
relativity in the IR region, the effective coupling is related to
the speed of light $c$, the Newton coupling $G$ and the effectively
cosmological constant $\Lambda $ as
\begin{equation}
\label{eq5} c = \frac{\mu \kappa ^2}{4}\sqrt {\frac{\Lambda _W }{1 -
3\lambda }} , \quad G_N = \frac{\kappa ^2c}{32\pi }, \quad \Lambda =
\frac{3}{2}\Lambda _W .
\end{equation}
From the action, the spherically symmetric black hole solution can
be obtained as

\begin{equation}
\label{eq6} ds^2 = - f\left( r \right)c^2dt^2 + \frac{1}{f\left( r
\right)}dr^2 + r^2d\Omega _k^2 ,
\end{equation}

\noindent with $f\left( r \right) = k + \frac{\tilde {\mu
}x^2}{2\tilde {\beta }}\left( {1 - \sqrt {1 - \frac{4\tilde {\beta
}}{x^2\tilde {\mu }^2}\left( {\tilde {\mu }x^2 - \sqrt {c_0 x} }
\right)} } \right)$, $\tilde {\mu } = - \mu \Lambda _W $, $\tilde
{\beta } = \frac{\Lambda _W^2 }{4M}$, $x = \sqrt { - \Lambda _W }
r$, $k = 0,\pm 1$, $\Lambda _W $ is the cosmological constant. The
mass of the black hole is expressed as $m = \frac{1}{16}k^2\Omega _k
c_0 \left( { - \Lambda _W } \right)^{ - \frac{3}{2}}$, $c_0 \ge 0$
is an integration constant and can be expressed in terms of horizon
radius as $c_0 = \frac{1}{x_ + }\left( {\frac{\tilde {\beta }k^2}{x_
+ ^2 } + \tilde {\mu }\left( {k + x_ + ^2 } \right)} \right)^2$ and
$d\Omega _k^2 $ is the line element for a two-dimensional Einstein
space with constant scalar curvature $2k$ and volume $\Omega _k $.
To have a well-defined vacuum solution, it should let $\frac{4\tilde
{\beta }}{\tilde {\mu }} \le 1$. The solution is asymptotically
$AdS_4 $. When $\tilde {\beta }$ goes to zero, we can get $f = k +
x^2 - \frac{1}{\tilde {\mu }}\sqrt {c_0 x} $ from the equations of
motion given in Ref. [11]. Meanwhile when an appropriate integration
constant is chosen and $k = 1$, the metric (6) reduces to the
solutions given in Ref. [6] and Ref. [5], respectively. In the
following, we investigate the Hawking radiation of spin 1/2
particles in Horava-Lifshitz gravitation theory. To explore the
fermion tunnelling, we introduce Dirac equation

\begin{equation}
\label{eq7} i\gamma ^\mu \left( {\partial _\mu + \Omega _\mu }
\right)\Psi + \frac{m}{\hbar }\Psi = 0,
\end{equation}

\noindent where $\Omega _\mu = \frac{i}{2}\Gamma _\mu ^{\alpha \beta
} \sum _{\alpha \beta } $, $\sum _{\alpha \beta } =
\frac{i}{4}\left[ {\gamma ^\alpha ,\gamma ^\beta } \right]$, $\gamma
^\mu $ matrices satisfy $\left\{ {\gamma ^\mu ,\gamma ^\nu }
\right\} = 2g^{\mu \nu }I$, and $m$ is the mass of the emission
fermion. For getting the solution of the Dirac equation, we first
choose $\gamma ^\mu $ matrices. There are many ways to choose them,
and our choice is given as

\[
\gamma ^t = \frac{1}{\sqrt {f\left( r \right)} }\left(
{{\begin{array}{*{20}c}
 0 \hfill & I \hfill \\
 { - I} \hfill & 0 \hfill \\
\end{array} }} \right),
\quad \gamma ^\theta = \sqrt {g^{\theta \theta }} \left(
{{\begin{array}{*{20}c}
 0 \hfill & {\sigma ^1} \hfill \\
 {\sigma ^1} \hfill & 0 \hfill \\
\end{array} }} \right),
\]

\begin{equation}
\label{eq8} \gamma ^r = \sqrt {f\left( r \right)} \left(
{{\begin{array}{*{20}c}
 0 \hfill & {\sigma ^3} \hfill \\
 {\sigma ^3} \hfill & 0 \hfill \\
\end{array} }} \right),
\quad \gamma ^\phi = \sqrt {g^{\phi \phi }} \left(
{{\begin{array}{*{20}c}
 0 \hfill & {\sigma ^2} \hfill \\
 {\sigma ^2} \hfill & 0 \hfill \\
\end{array} }} \right),
\end{equation}
$\sigma ^\mu $ is the Pauli sigma matrix. For a spin 1/2 particle,
there are two states, which correspond to spin up state and spin
down state, respectively. When we measure the spin states along $r$
direction, the spin up case is along $r$ direction and the spin down
case has opposite direction. In this Letter, we only investigate the
spin up case. The spin down case is similar as that of spin up state
and the same result can be gotten [51]. The wave function of the
spin up case is given by
\begin{equation}
\label{eq9} \psi _{\left( \uparrow \right)} = \left(
{{\begin{array}{*{20}c}
 {A\left( {t,r,\theta ,\phi } \right)} \hfill \\
 0 \hfill \\
 {B\left( {t,r,\theta ,\phi } \right)} \hfill \\
 0 \hfill \\
\end{array} }} \right)\exp \left( {\frac{i}{\hbar }I_ \uparrow \left(
{t,r,\theta ,\phi } \right)} \right).
\end{equation}
Inserting the wave function and  $\gamma ^\mu $ matrices into the
Dirac equation, dividing the exponential term and multiplying by
$\hbar $, we can get the resulting equations to leading order in
$\hbar $ as

\begin{equation}
\label{eq10} \frac{B}{\sqrt {f\left( r \right)} }\partial _t I_
\uparrow + B\sqrt {f\left( r \right)} \partial _r I_ \uparrow - mA =
0,
\end{equation}

\begin{equation}
\label{eq11} \frac{ - A}{\sqrt {f\left( r \right)} }\partial _t I_
\uparrow + A\sqrt {f\left( r \right)} \partial _r I_ \uparrow - mB =
0,
\end{equation}

\begin{equation}
\label{eq12} B\sqrt {g^{\theta \theta }} \partial _\theta I_
\uparrow + iB\sqrt {g^{\phi \phi }} \partial _\phi I_ \uparrow = 0,
\end{equation}

\begin{equation}
\label{eq13} A\sqrt {g^{\theta \theta }} \partial _\theta I_
\uparrow + iA\sqrt {g^{\phi \phi }} \partial _\phi I_ \uparrow = 0.
\end{equation}
It is difficult to solve the action of the emission fermion. From
above equations, we know the action can be separated into radial
term and transverse term. Considering the symmetry of the spacetime,
we carry out the separation of variables as

\begin{equation}
\label{eq14} I_ \uparrow = - \omega t + W\left( r \right) + \Theta
\left( {\theta ,\phi } \right),
\end{equation}

\noindent where $\omega $ is the energy of the particle. The
contribution of imaginary part of the action is mainly produced by
two parts, namely $ W\left( r \right) $ and $\Theta \left( {\theta
,\phi } \right)$, but the contribution of $\Theta \left( {\theta
,\phi } \right)$ could be canceled out in the calculation of the
tunnelling probability. So we only take care of the action of the
radial direction here. Inserting Eq. (14) into (10) and (11) yields

\begin{equation}
\label{eq15}
 - \frac{B\omega }{\sqrt {f\left( r \right)} } + B\sqrt {f\left( r \right)}
\partial _r W\left( r \right) - mA = 0,
\end{equation}

\begin{equation}
\label{eq16} \frac{A\omega }{\sqrt {f\left( r \right)} } + A\sqrt
{f\left( r \right)}
\partial _r W\left( r \right) - mB = 0.
\end{equation}
When $m = 0$, it denotes the Hawking radiation of the massless
particle. While $m \ne 0$, it is the Hawking radiation of the
massive fermion. The final results (tunnelling probability and
Hawking temperature) are not related to the mass of the particle.
Without loss the generality in the discussion, we choose $m \ne 0$.
Solving the action of radial direction and canceling out A and B, we
get

\[
W_\pm \left( r \right) = \pm \int {\frac{\sqrt {\omega ^2 +
m^2f\left( r \right)} }{f\left( r \right)}dr} = \pm \frac{i\pi
\omega }{{f}'\left( {r_ + } \right)},
\]

\begin{equation}
\label{eq17} {f}'\left( {r_ + } \right) = \frac{\sqrt { - \Lambda _W
} }{2x_ + }\frac{3x_ + ^2 - k - \frac{5\tilde {\beta }k^2}{\tilde
{\mu }x_ + ^2 }}{1 + \frac{2\tilde {\beta }k}{\tilde {\mu }x_ + ^2
}},
\end{equation}

\noindent where $ + / - $ correspond to the outgoing/ingoing
solutions. From the WKB approximation, we know the tunnelling
probability is related to the imaginary part of the action. Thus the
tunnelling probability of the emission fermion is obtained as

\[
\Gamma = \frac{P\left( {emission} \right)}{P\left( {absorption}
\right)} = \frac{\exp \left( { - 2ImI_ + } \right)}{\exp \left( { -
2ImI_ - } \right)} = \frac{\exp \left( { - 2ImW_ + } \right)}{\exp
\left( { - 2ImW_ - } \right)}
\]

\begin{equation}
\label{eq18}
 = \exp \left[ { - \frac{4\pi \omega }{{f}'\left( {x_ + } \right)}}
\right],
\end{equation}

\noindent which means the Hawking temperature is
\begin{equation}
\label{eq19} T = \frac{{f}'\left( {x_ + } \right)}{4\pi } =
\frac{\sqrt { - \Lambda _W } }{8\pi x_ + }\frac{3x_ + ^2 - k -
\frac{5\tilde {\beta }k^2}{\tilde {\mu }x_ + ^2 }}{1 + \frac{2\tilde
{\beta }k}{\tilde {\mu }x_ + ^2 }},
\end{equation}

\noindent where $x_ + $ is the black hole horizon obtained from
$f\left( r \right) = 0$. When $\tilde {\beta } \to 0$, it is just
the Hawking temperature of the black hole in Horava-Lifshitz gravity
with $z = 3$ [6] and recovers the Hawking temperature of the black
hole in Horava-Lifshitz gravity where $k = 1$ that was obtained by
L\"{u} [5]. When $k = 0$, the Hawking temperature is  $T = \frac{3x_
+ \sqrt { - \Lambda _W } }{8\pi }$. In the extreme case, when $k \ne
1$ and $\frac{\tilde {\beta }}{\tilde {\mu }} = \frac{x_ + ^2 \left(
{3x_ + ^2 - k} \right)}{5}$, the Hawking temperature vanishes. The
Hawking temperature vanishing means the surface gravity of the black
hole vanishes, but the entropy of the black hole doesn't vanish.
When $k = 1$ and $\frac{\tilde {\beta }}{\tilde {\mu }} > \frac{x_ +
^2 \left( {3x_ + ^2 - 1} \right)}{5}$, the temperature is negative.
When $k = - 1$ and $\frac{x_ + ^2 }{2} > \frac{\tilde {\beta
}}{\tilde {\mu }} > \frac{x_ + ^2 \left( {3x_ + ^2 + 1}
\right)}{5}$, the temperature is also negative. (Note the condition
$4\frac{\tilde {\beta }}{\tilde {\mu }} \le 1$ of the vacuum
solution is considered here [11]). The negative temperature is not
allowed in black hole physics. Therefore we have to avoid the
negative temperature. When we consider $\tilde {\mu } = - \mu
\Lambda _W $, $\tilde {\beta } = \frac{\Lambda _W^2 }{4M}$  and the
expression of the black hole horizon $x_+$, we find the Hawking
temperatures are related to the fundamental parameters ($\mu$, $M$,
$\Lambda _W$). Therefore we must restrict the fundamental parameters
of Horava-Lifshitz gravity to avoid negative temperatures. This is
an important constraint on the theory.

The Hawking temperatures change with the parameters is expressed in
the following figures. There both $c_0 $ and $\Lambda _W $ are
choose to equal 1 for visibility. From the Figure 1 and Figure 2, we find the
temperatures become lower with the increase of $M$ (or $\mu )$. When
$M$ (or $\mu )$ is fixed, the bigger $\mu $ ( or $M$) corresponds
the lower temperatures.

\begin{figure}[htb]
\includegraphics[width=0.45\textwidth]{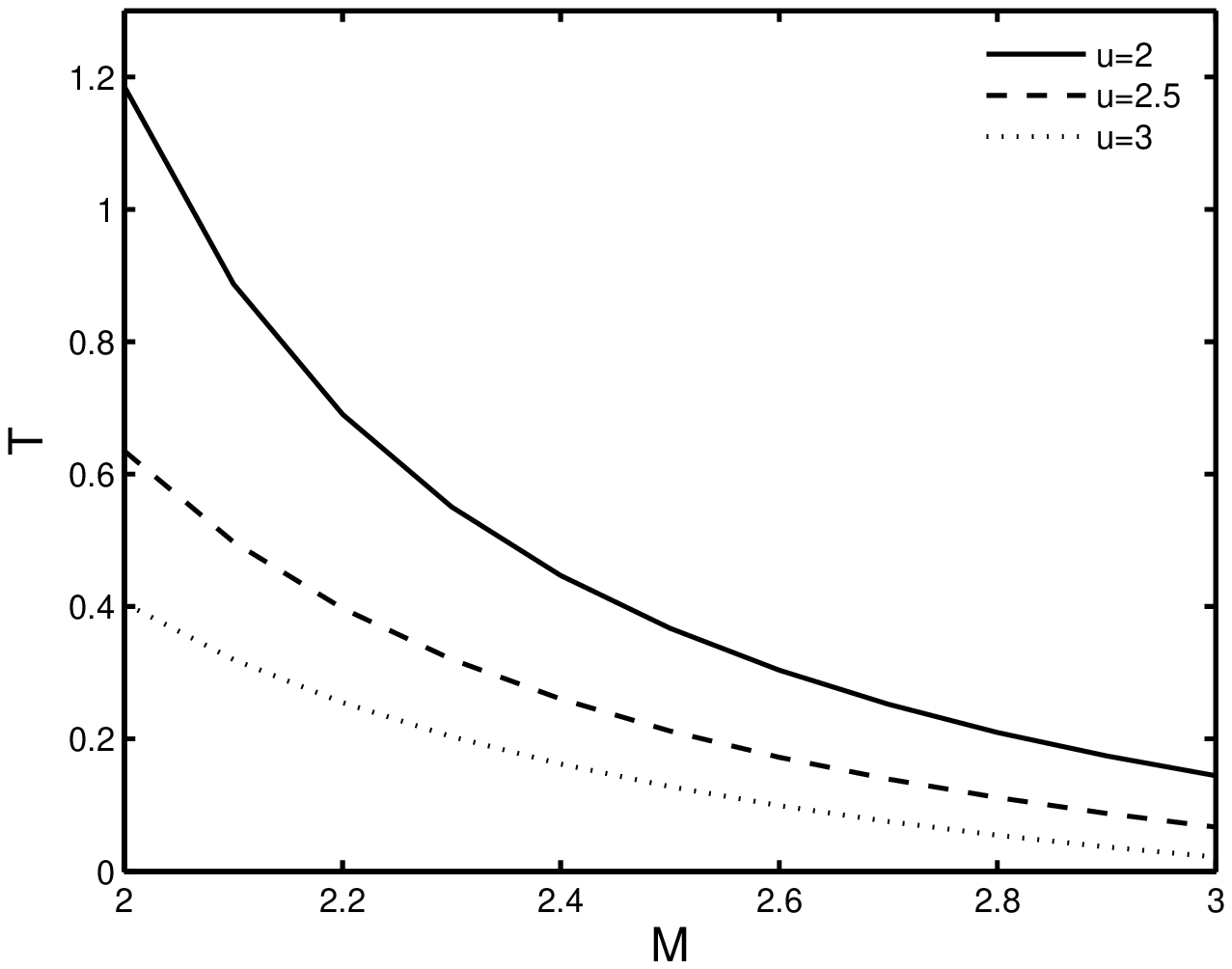}
\vspace{-0.25cm} \includegraphics[width=0.45\textwidth]{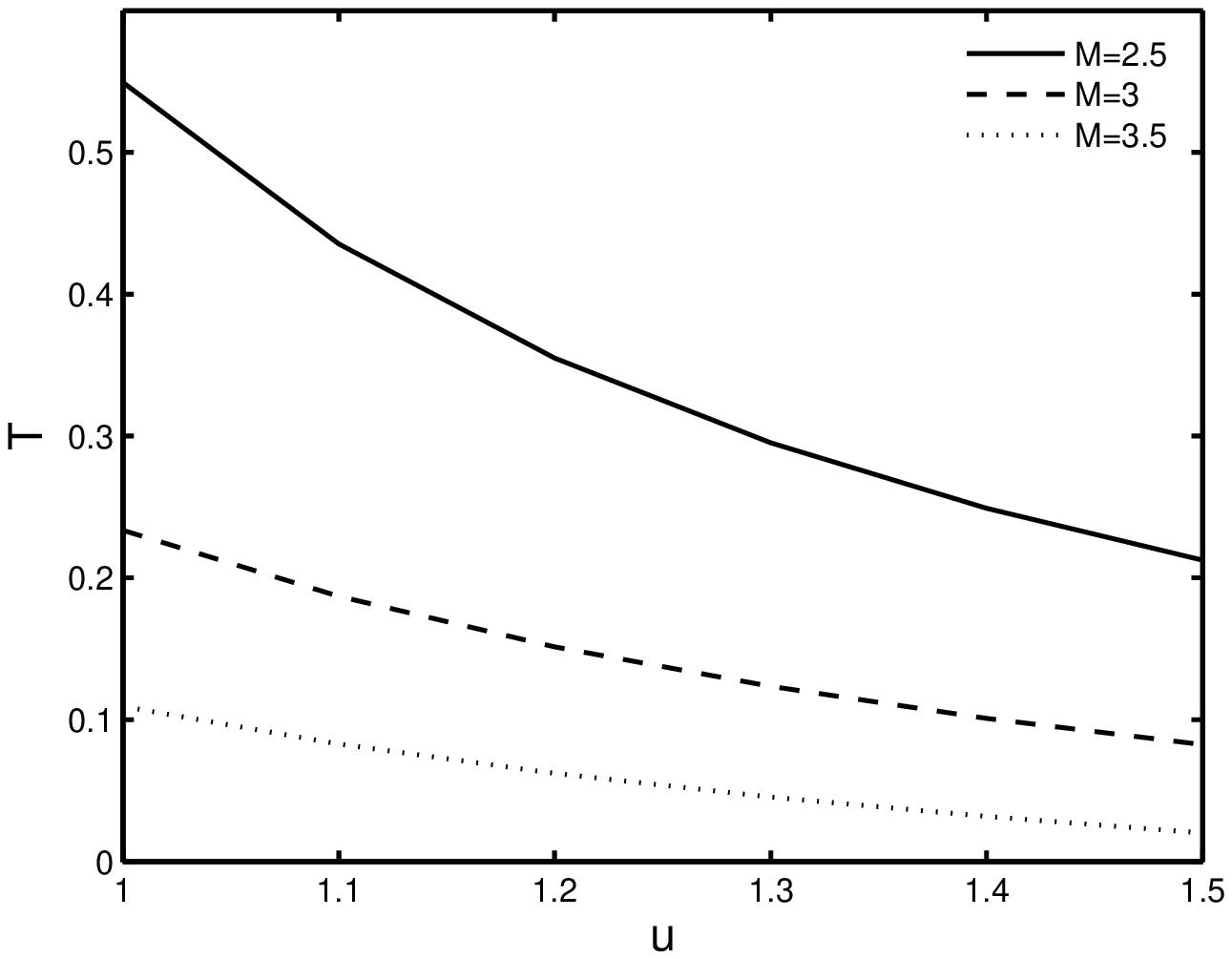}
\vspace{-0.25cm}\caption{The Hawking temperature for the case of
$k=1$} \label{fig:bosons}
\end{figure}

\begin{figure}[htb]
\includegraphics[width=0.45\textwidth]{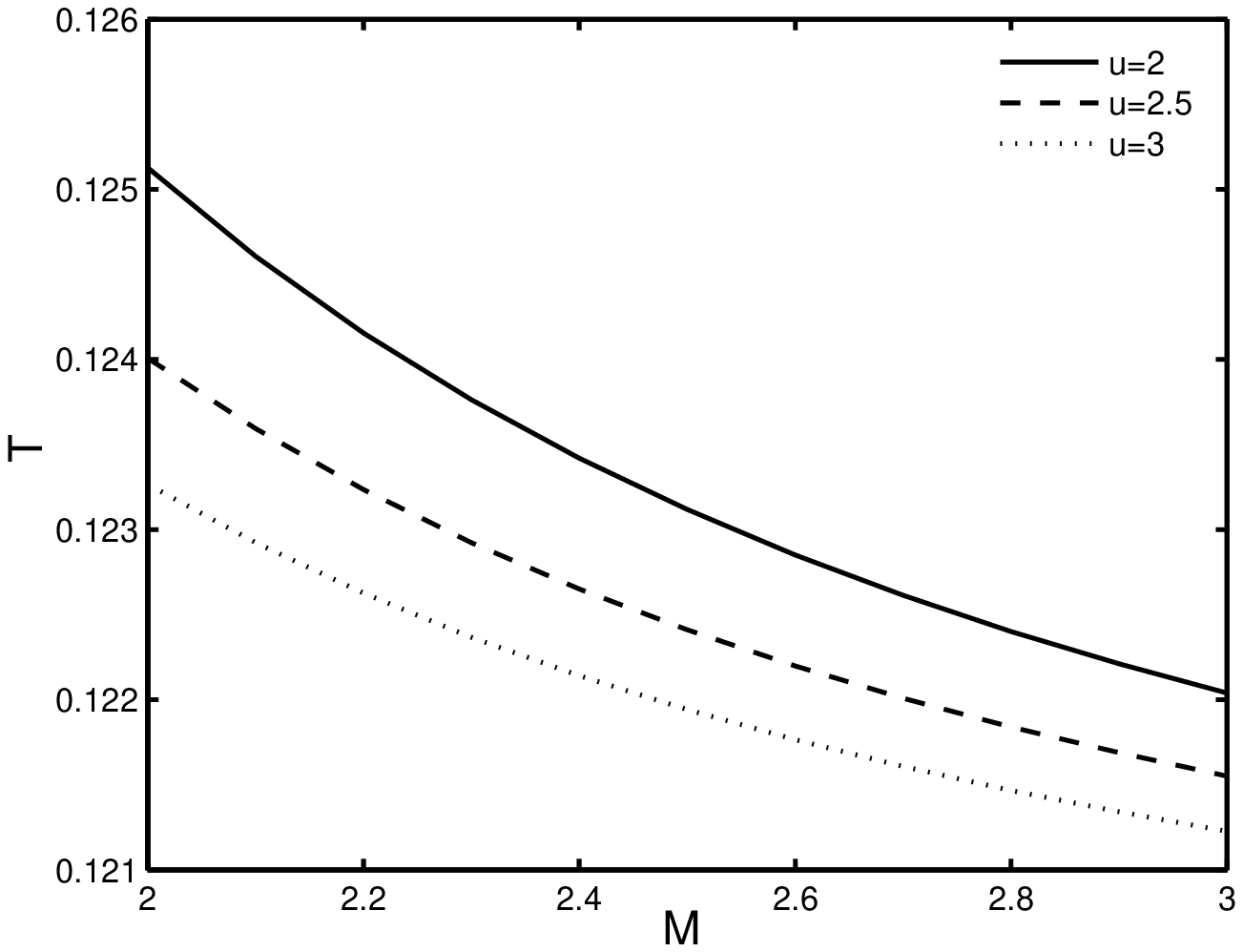}
\vspace{-0.25cm}\includegraphics[width=0.45\textwidth]{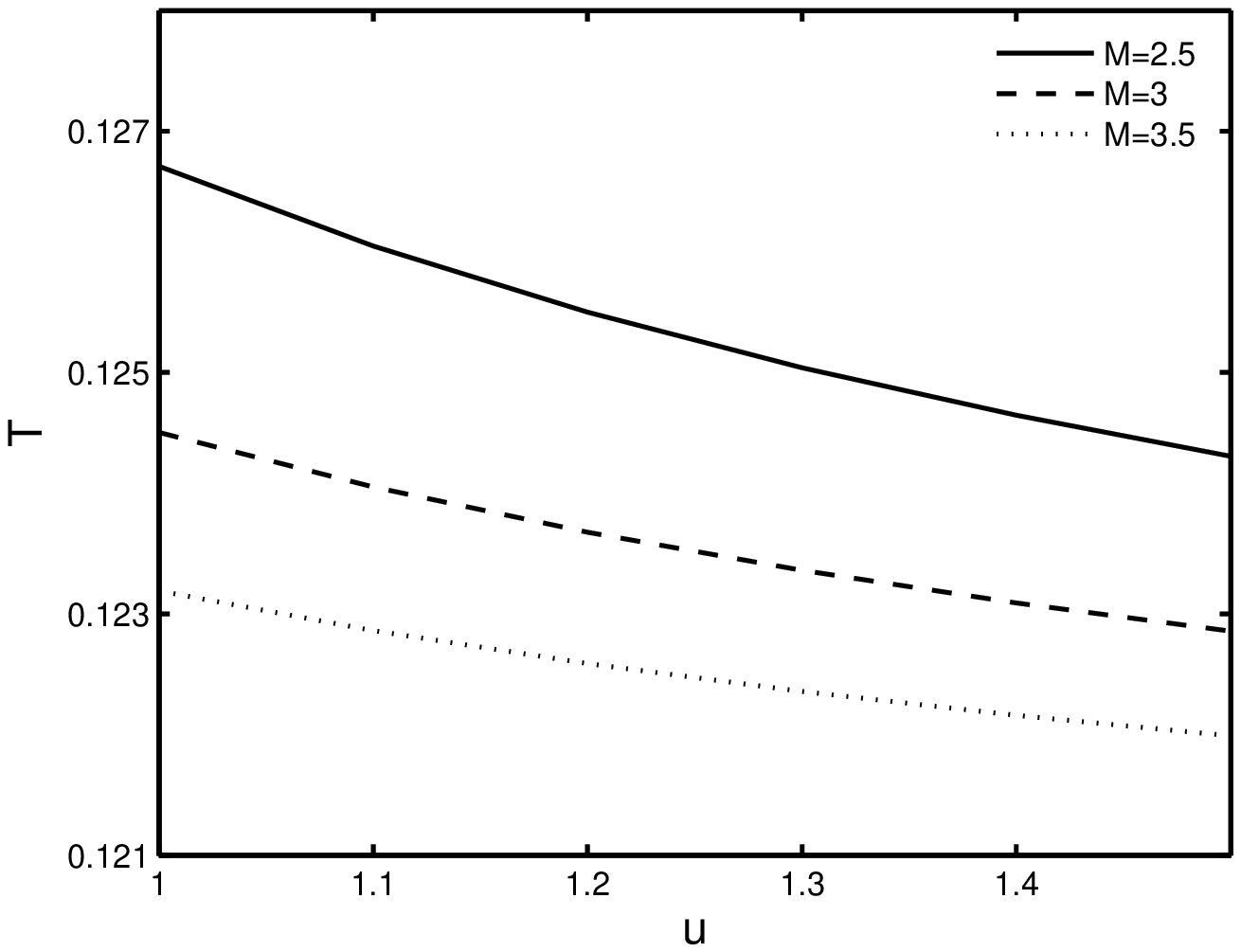}
\vspace{-0.25cm}\caption{The Hawking temperature for the case of
$k=-1$} \label{fig:bosons}
\end{figure}

In the calculation, we only investigated the Hawking radiation of
the spin up state. For the spin down case, we can adopt analogous
step to investigate it and get the same result. In the
investigation, the background spacetime of black holes was seen as
fixed with the particle emission and the back reaction is neglected.
The true tunnelling picture implies the background should fluctuate
with particles emission. Therefore considering the true tunnelling
picture, the Hawking temperature should be corrected. Meanwhile we
neglected the higher order terms of the action in the calculation.
When it was considered, the Hawking temperature would be also
corrected. This view has been mentioned in Ref. [54, 56], and we
don't correct the temperature in the current Letter.

\textbf{3. Hawking radiation of (4 + 1)-dimensional black holes in
the $z = 4$ Horava-Lifshitz gravity}

In this section, we investigate the fermions tunnelling effect in
(4+1)-dimensional spacetime in the $z = 4$ Horava-Lifshitz gravity.
This may be interesting to study the $AdS_5 / CFT_4 $ correspondence
in the framework of this theory.

The Lagrangian of the $z = 4$ Horava-Lifshitz gravity in
(4+1)-dimensional spacetime is given by

\[\mathcal {L} = \mathcal {L}_0 + \mathcal {L}_1,\]

\[
\mathcal {L}_0 = \sqrt g N\left\{ {\frac{2}{\kappa ^2}\left(
{K^{ij}K{ }_{ij} - \lambda K^2} \right) + \frac{\kappa ^2\mu
^2\Lambda _W \left( {R - 2\Lambda _W } \right)}{4\left( {1 -
4\lambda } \right)}} \right\},
\]

\[
\mathcal {L}_1 = - \sqrt g N\frac{\kappa ^2}{8}\left\{ {\mu ^2G_{ij}
G^{ij} + \frac{2\mu }{M}G^{ij}L_{ij} + \frac{2\mu }{M}\Lambda _W L +
\frac{1}{M^2}L^{ij}L_{ij} } \right.
\]

\begin{equation}
\label{eq20} \left. { - \tilde {\lambda }\left( {\frac{L^2}{M^2} -
\frac{2\mu L}{M}\left( {R - 4\Lambda _W } \right) + \mu ^2R^2}
\right)} \right\} ,
\end{equation}

\noindent where $L = 2\left( {1 + 3\beta } \right)\nabla ^2R$, the
expressions of $L^{ij}$ and $G^{ij}$ are given in section 2. In
order to get back to general relativity in the IR region, the
effective couplings should be related to the light speed, the Newton
coupling and the effectively cosmological constant as

\begin{equation}
\label{eq21} c = \frac{\mu \kappa ^2}{\sqrt 8 }\sqrt {\frac{\Lambda
_W }{1 - 3\lambda }} , \quad G_N = \frac{\kappa ^2c}{32\pi }, \quad
\Lambda = \Lambda _W .
\end{equation}
In the IR region, to get back to general reactivity, we let $\lambda
$ be equal to 1, and then $\Lambda _W $ should be negative for the
speed light having the physical meaning. It is difficult to get the
exact solutions for the general $\beta $. In this section, the value
of $\beta $ is chosen as -1/3, and it comes from an unsuccessful
attempt to generalize the New Massive Gravity to the case in four
dimensions. From the action, the solution of (4+1)-dimensional black
holes solution in Horava-Lifshitz gravity was given by [11]

\begin{equation}
\label{eq22} ds^2 = - F\left( r \right)c^2dt^2 + \frac{1}{F\left( r
\right)}dr^2 + r^2d\Omega _k^2 ,
\end{equation}

\noindent where $F\left( r \right) = k + \frac{x^2}{3}\pm \sqrt {c_0
} $, $c_0 \ge 0$ is an integration constant, $d\Omega _k^2 $ is the
three-dimensional Einstein manifold with constant scalar curvature
$6k$, which one can choose $k = 0,\pm 1$, without loss of
generality. The metric (22) denotes a five-dimensional Anti de
Sitter spacetime obtained from the Horava-Lifshitz gravity. It can
not reduce to the 5-dimensional Schwarzschild metric. The quantum
inheritance principle is an important principle in the construction
of the Horava-Lifshitz gravity. If the Horava-Lifshitz gravity is
quantum gravity, it is very important to understand the AdS/CFT
correspondence in the frame work of Horava-Lifshitz theory and it is
interested to further study the asymptotically AdS solution in
five-dimensional spacetime. Therefore we further study the Hawking
radiation of fermions from these black holes in this section.

To investigate the Hawking radiation of emission fermions, we still
use the Dirac equation expressed in Eq. (7) and spin up wave
function (9). For the five-dimensional spacetime, we choose the
following gamma matrices

\[
\gamma ^t = \frac{1}{\sqrt {F\left( r \right)} }\left(
{{\begin{array}{*{20}c}
 i \hfill & 0 \hfill \\
 0 \hfill & { - i} \hfill \\
\end{array} }} \right),
\quad \gamma ^r = \sqrt {F\left( r \right)} \left(
{{\begin{array}{*{20}c}
 0 \hfill & {\sigma ^3} \hfill \\
 {\sigma ^3} \hfill & 0 \hfill \\
\end{array} }} \right),
\quad \gamma ^\theta = \sqrt {g^{\theta \theta }} \left(
{{\begin{array}{*{20}c}
 0 \hfill & {\sigma ^2} \hfill \\
 {\sigma ^2} \hfill & 0 \hfill \\
\end{array} }} \right),
\]

\begin{equation}
\label{eq23} \gamma ^\varphi = \sqrt {g^{\varphi \varphi }} \left(
{{\begin{array}{*{20}c}
 0 \hfill & {\sigma ^1} \hfill \\
 {\sigma ^1} \hfill & 0 \hfill \\
\end{array} }} \right),
\quad \gamma ^\phi = \sqrt {g^{\phi \phi }} \left(
{{\begin{array}{*{20}c}
 { - I} \hfill & 0 \hfill \\
 0 \hfill & I \hfill \\
\end{array} }} \right).
\end{equation}
Inserting the spin up wave function and gamma matrices into the
Dirac equation, dividing the exponential term and multiplying by
$\hbar $, we can get the resulting equations to leading order in
$\hbar $ as
\begin{equation}
\label{eq24} \frac{iA}{\sqrt {F\left( r \right)} }\partial _t I_
\uparrow + B\sqrt {F\left( r \right)} \partial _r I_ \uparrow -
A\sqrt {g^{\phi \phi }}
\partial _\phi I_ \uparrow - mA = 0,
\end{equation}

\begin{equation}
\label{eq25} \frac{ - iB}{\sqrt {F\left( r \right)} }\partial _t I_
\uparrow + A\sqrt {F\left( r \right)} \partial _r I_ \uparrow +
B\sqrt {g^{\phi \phi }}
\partial _\phi I_ \uparrow - mB = 0,
\end{equation}

\begin{equation}
\label{eq26} iB\sqrt {g^{\theta \theta }} \partial _\theta I_
\uparrow + B\sqrt {g^{\varphi \varphi }} \partial _\varphi I_
\uparrow = 0,
\end{equation}

\begin{equation}
\label{eq27} iA\sqrt {g^{\theta \theta }} \partial _\theta I_
\uparrow + A\sqrt {g^{\varphi \varphi }} \partial _\varphi I_
\uparrow = 0.
\end{equation}
It is difficult to solve the above equations. Similar to that in the
section 2, considering the symmetry of the spacetime, we carry out
the separation of variable as

\begin{equation}
\label{eq28} I_ \uparrow = - \omega t + W\left( r \right) + \Xi
\left( {\theta ,\varphi ,\phi } \right),
\end{equation}
Insert Eq. (28) into (24)-(27). Considering the contribution of the
imaginary part of $ \Xi \left( {\theta ,\varphi ,\phi } \right)$
would be canceled out in the calculation of the tunnelling
probability, we still only take care of the radial direction part.
Solving $W\left( r \right)$ yields
\[
W_\pm \left( r \right) = \pm \int {\frac{\sqrt {\omega ^2 + \left(
{\sqrt {g^{\phi \phi }} \partial _\phi \Xi \left( {\theta ,\varphi
,\phi } \right) - m^2 - \frac{2m\omega }{\sqrt {F\left( r \right)}
}} \right)F\left( r \right)} }{F\left( r \right)}dr}
\]

\begin{equation}
\label{eq29}
 = \pm \frac{i\pi \omega }{{F}'\left( r \right)} = \pm \frac{i3\pi \omega
}{2x_ + \sqrt { - \Lambda _W } }.
\end{equation}

\noindent where $ + / - $ correspond to the outgoing/ingoing
solutions. Applying the WKB approximation, we obtain the tunnelling
probability of the emission fermion in (4+1)-dimensional spacetime
as

\[
\Gamma = \frac{P\left( {emission} \right)}{P\left( {absorption}
\right)} = \frac{\exp \left( { - 2ImI_ + } \right)}{\exp \left( { -
2ImI_ - } \right)} = \frac{\exp \left( { - 2ImW_ + } \right)}{\exp
\left( { - 2ImW_ - } \right)}
\]

\begin{equation}
\label{eq30}
 = \exp \left( { - \frac{6\pi \omega }{\sqrt { - \Lambda _W } x_ + }}
\right),
\end{equation}

\noindent which results the Hawking temperature as

\begin{equation}
\label{eq31} T = \frac{\sqrt { - \Lambda _W } x_ + }{6\pi }.
\end{equation}
This result is in consistence with that obtained by directly
calculating surface gravity of the black holes. For the spin down
case, we can use analogous process to explore the Hawking radiation
of the black holes and get the same temperature. The unfixed
background spacetime and back reaction were also neglected in this
section. When these are considered, the Hawking temperature should
be corrected.

\textbf{4. Discussion and Conclusion}

In this Letter, we have investigated the Hawking radiation of (3+1)-
and (4+1)-dimensional black holes in the $z = 4$ Horava-Lifshitz
gravity by fermion tunnelling. As a result, the Hawking temperatures
were recovered and are in consistence with that obtained by directly
calculating surface gravity of the black holes. The result shows the
Hawking temperatures of (3+1)-dimensional black holes in the $z=4$
Horava-Lifshitz gravity are related to the fundamental parameters
($\mu$, $M$, $\Lambda _W$) of Horava-Lifshitz gravity. Considering
the negative temperature is not allowed in black hole physics,
therefore we have to restrict the fundamental parameters to avoid
the negative temperatures. For the (4+1)-dimensional black hole, the
Hawking temperatures are related to the cosmological constant. This
is our expectation. In the investigation, the unfixed background
spacetime and back reaction were neglected, and then the obtained
Hawking temperatures are only leading term. When they are taken into
account, the Hawking temperatures should be corrected. Recently
Banerjee and Majhi et. al. made a correction to the Hawking
temperature by Hamilton-Jacobi method. They extended the action of
the emission particle in a power of $\hbar $ and made a correction
to the action, and then obtained the correction temperatures. When
the action of emission particle is corrected in this Letter, the
corrected temperatures can also be obtained.

\textbf{Acknowledgments}

This work is supported by the Natural Science Found of China under
Grant Nos. 10705008, 10773008.

\textbf{References.}

[1] P. Horava, Quantum Criticality and Yang-Mills Gauge Theory,
arXiv: 0811.2217 [hep-th].

[2] P. Horava, JHEP 0903 (2009) 020.

[3] P. Horava, Phys. Rev. D79 (2009) 084008.

[4] P. Horava, Phys. Rev. Lett. 102 (2009) 161301.

[5] H. L\"{u}, J.W. Mei and C.N. Pope, Solution to Horava Gravity.
arXiv: 0904.1595 [hep-th].

[6] R.G. Cai, Y. Liu and Y.W. Sun, On the z=4 Horava-Lifshitz
Gravity. arXiv:0904.4104 [hep-th]

[7] R.G. Cai, M.L. Cao and N. Ohta, Topological Black Holes in
Horava-Lifshitz Gravity. arXiv:0904.3670 [hep-th].

[8] A. Kehagias and K. Sfetsos, The black hole and FRW geometries of
non-relativistic gravity, arXiv:0905.0477 [hep-th].

[9] Eoin O. Colgain and H. Yavartanoo, Dyonic solution of
Horava-Lifshitz Gravity, arXiv:0904.4357 [hep-th].

[10] A. Ghodsi, Toroidal solutions in Horava Gravity,
arXiv:0905.0836 [hep-th];

A. Ghodsi and E. Hatefi, Extremal rotating solutions in Horava
Gravity, arXiv:0906.1237 [hep-th].

[11] M. Park, The Black Hole and Cosmological Solutions in IR
modified Horava Gravity, arXiv:0905.4480 [hep-th].

[12] T. Takahashi and J. Soda, Chiral Primordial Gravitational Waves
from a Lifshitz Point, arXiv:0904.0554 [hep-th].

[13] G. Calcagni, Cosmology of the Lifshitz universe,
arXiv:0904.0829 [hep-th].

[14] E. Kiritsis and G. Kofinas, Horava-Lifshitz Cosmology,
arXiv:0904.1334 [hep-th].

[15] R. Brandenberger, Matter Bounce in Horava-Lifshitz Cosmology,
arXiv:0904.2835 [hep-th].

[16] B. Chen, S. Pi and J. Z. Tang, Scale Invariant Power Spectrum
in Horava-Lifshitz Cosmology without Matter, arXiv:0905.2300
[hep-th].

[17] S. Nojiri and Sergei D. Odintsov, Covariant Horava-like
renormalizable gravity and its FRW cosmology, arXiv:0905.4213
[hep-th].

[18] S. Mukohyama, Scale-invariant cosmological perturbations from
Horava-Lifshitz gravity without inflation, arXiv:0904.2190 [hep-th].

[19] Y.S. Piao, Primordial Perturbation in Horava-Lifshitz
Cosmology, arXiv:0904.4117 [hep-th].

[20] X. Gao, Cosmological Perturbations and Non-Gaussianities in
Horava-Lifshitz Gravity, arXiv:0904.4187 [hep-th];

X. Gao, Y. Wang, R. Brandenberger and A. Riotto, Cosmological
Perturbations in Horava-Lifshitz Gravity, arXiv:0905.3821 [hep-th].

[21] A.Z. Wang, D. Wands and R. Maartens, Scalar field perturbations
in Horava-Lifshitz cosmology,  arXiv:0909.5167 [hep-th];

A.Z Wang and R. Maartens, Cosmological perturbations in
Horava-Lifshitz theory without detailed balance, arXiv:0907.1748
[hep-th];

A.Z. Wang and Y.M. Wu, Thermodynamics and classification of
cosmological models in the Horava-Lifshitz theory of gravity,
arXiv:0905.4117 [hep-th].

[22] R.G. Cai, M.L. Cao and N. Ohta, Thermodynamics of Black Holes
in Horava-Lifshitz Gravity. arXiv:0905.0751 [hep-th].

[23] Y.S. Myung and Y.W. Kim, Thermodynamics of Horava-Lifshitz
black holes, arXiv:0905.0179 [hep-th];

Y.S. Myung, Thermodynamics of black holes in the deformed
Horava-Lifshitz gravity, arXiv:0905.0957 [gr-qc]].

[24] D. Blas, O. Pujolas, S. Sibiryakov, JHEP 0910 (2009) 029
[arXiv: 0906.3406 [hep-th]].

[25] S.B. Chen and J.L. Jing, Quasinormal modes of a black hole in
the deformed Horava-Lifshitz gravity, arXiv:0905.1409 [gr-qc];
Strong field gravitational lensing in the deformed Horava-Lifshitz
black hole, arXiv:0905.2055 [gr-qc].

[26] J.J. Peng and S.Q. Wu, Hawking Radiation of Black Holes in
Infrared Modified Horava-Lifshitz Gravity, arXiv:0906.5121 [hep-th].

[27] E.N. Saridakis, Horava-Lifshitz Dark Energy, arXiv:0905.3532
[hep-th];

C. Bogdanos and E.N. Saridakis, Perturbative instabilities in Horava
gravity,  arXiv:0907.1636 [hep-th];

Y.F. Cai and E.N. Saridakis, Non-singular cosmology in a model of
non-relativistic gravity, arXiv:0906.1789 [hep-th];

G. Leon and E.N. Saridakis, Phase-space analysis of Horava-Lifshitz
cosmology, arXiv:0909.3571 [hep-th].

[28] J. Kluson, Brane at quantum criticality, arXiv:0904.1343
[hep-th].

[29] B. Chen and Q.G. Huang, Field Theory at a Lifshitz Point,
arXiv: 0904.4565 [hep-th].

[30] T. Nishioka, Horava-Lifshitz Holography, arXiv:0905.0473
[hep-th].

[31] M. Li and Y. Pang, A Trouble with Horava-Lifshitz Gravity,
arXiv: 0905.2751 [hep-th].

[32] T.P. Sotiriou, M. Visser and S. Weinfurtner, Quantum gravity
without Lorentz invariance, arXiv:0905.2798 [hep-th].

[33] H. Nastase, On IR solutions in Horava gravity theories,
arXiv:0904.3604 [hep-th].

[34] H. Nikolic, Horava-Lifshitz gravity, absolute time, and
objective particles in curved space, arXiv:0904.3412 [hep-th].

[35] G.E. Volovik, z=3 Lifshitz-Horava model and Fermi-point
scenario of emergent gravity, arXiv:0904.4113 [gr-qc].

[36] R.G. Cai, B. Hu and H.B. Zhang, Dynamical Scalar Degree of
Freedom in Horava-Lifshitz Gravity, arXiv:0905.0255 [hep-th].

[37] J.H. Chen and Y.J. Wang, Timelike Geodesic Motion in
Horava-Lifshitz Spacetime, arXiv:0905.2786 [gr-qc].

[38] D. Orlando and S. Reffert, On the Renormalizability of
Horava-Lifshitz-type Gravities, arXiv: 0905.0301 [hep-th].

[39] C. Charmousis, G. Niz, A. Padilla and Paul M. Saffin, JHEP 0908
(2009) 070 [arXiv:0905.2579 [hep-th]].

[40] M.K. Parikh and F. Wilczek, Phys. Rev. Lett. 85 (2000) 5024.

[41] S.P. Robinson and F. Wilczek, Phys. Rev. Lett. 95 (2005)
011303.

[42] R. Banerjee and S. Kulkarni, Phys. Rev. D77 (2008) 024018
[arXiv: 0707.2449 [hep-th]]; Phys. Lett. B659 (2008) 827
[arXiv:0709.3916 [hep-th]]; Phys. Rev. D79 (2009) 084035
arXiv:0810.5683 [hep-th];

R. Banerjee, Int. J. Mod. Phys. D17 (2009) 2539, [arXiv:0807.4637
[hep-th]];

R.Banerjee and B. R. Majhi, Phys. Lett. B662 (2008) 62
[arXiv:0801.0200 [hep-th]].

[43] K. Srinivasan and T. Padmanabhan, Phys. Rev. D 60 (1999) 024007
[arXiv:9812028 [gr-qc]];

S. Shankaranarayanan, T. Padmanabhan and K. Srinivasan, Hawking
radiation in different coordinate settings: Complex paths approach,
arXiv: 0010042[gr-qc];

S. Shankaranarayanan, K. Srinivasan and T. Padmanabhan, Mod. Phys.
Lett. A 16 (2001) 571;

M. Angheben, M. Nadalini, L. Vanzo and S. Zerbini, JHEP 05 (2005)
014;

R. Kerner, R.B. Mann, Phys. Rev. D73 (2006) 104010 [arXiv:0603019
[gr-qc]];

D.Y. Chen and S.Z. Yang, Gen. Rel. Grav. 39 (2007) 1503.

[44] S. Iso, H. Umetsu and F. Wilczek, Phys. Rev. Lett. 96 (2006)
151302;

K. Murata and J. Soda, Phys. Rev. D74 (2006) 044018.

[45] E.C. Vagenas and S. Das, JHEP 0610 (2006) 025.

[46] Z.B. Xu and B. Chen, Phys. Rev. D 75 (2007) 024041.

[47] E.T. Akhmedov, V. Akhmedova, T. Pilling and D. Singleton, Int.
J. Mod. Phys.A22 (2007) 1705 [arXiv:hep-th/0605137];

E.T. Akhmedov, V. Akhmedova and D. Singleton, Phys. Lett. B642
(2006) 124 [arXiv:hep-th/0608098]; Int. J. Mod. Phys. D17 (2008)
2453 [arXiv:0805.2653 [gr-qc]];

B.D. Chowdhury, Problems with Tunneling of Thin Shells from Black
Holes, arXiv: 0605197 [hep-th];

[48] Q.Q. Jiang and S.Q. Wu, Phys. Lett. B635 (2006) 151; Phys.
Lett. B647 (2007) 200;

Q.Q. Jiang, S.Q. Wu and X. Cai, Phys. Rev. D 73 (2006) 064003. Phys.
Rev. D75 (2007) 064029; Phys. Lett. B651 (2007) 58; Phys. Lett. B651
(2007) 65.

[49] S.Q. Wu and J.J. Peng, Class. Quant. Grav. 24 (2007) 5123;

S.Q. Wu, J.J. Peng and Z.Y Zhao, Class. Quant. Grav. 25 (2008)
135001;

S.Q. Wu and Z.Y Zhao, Comment on "Hawking Radiation and Covariant
Anomalies", arXiv: 0709.4074 [hep-th].

[50] J.Y. Zhang and Z. Zhao, Phys. Lett. B 618 (2005) 14; JHEP 0505
(2005) 055; Nucl. Phys. B725 (2005) 173.

[51] R. Kerner and R.B. Mann, Class. Quant. Grav. 25 (2008) 095014;
Phys. Lett. B665 (2008) 277.

[52] R. Di Criscienzo and L. Vanzo, Eur. phys. Lett. 82 (2008)
60001.

[53] R. Li and J.R. Ren, Class. Quant. Grav. 25 (2008) 125016; Phys.
Lett. B661 (2008) 370.

[54] R. Banerjee and S.K. Modak, JHEP 0905 (2009) 063
[arXiv:0903.3321];

R. Banerjee and B.R. Majhi, JHEP 0806 (2008) 095; Phys. Lett. B674
(2009) 218; Phys. Lett. B675 (2009) 243.

[55] S.A. Hayward, R. Di Criscienzo, M. Nadalini, L. Vanzo and S.
Zerbini, Local Hawking temperature for dynamical black holes,
arXiv:0806.0014 [hep-th].

[56] B.R. Majhi, Phys. Rev. D79 (2009) 044005.

[57] B. Chatterjee and P. Mitra, Phys. Lett. B675 (2009) 240.

[58] S. Bhattacharya and A. Saha, Scalar and spinor emissions from
G\"{o}del black hole, arXiv:0904.3441 [hep-th].

[59] T. Zhu, J.R. Ren, M.F. Li, Corrected Entropy of
Friedman-Robertson-Walker Universe in Tunneling Method,
arXiv:0905.1838 [hep-th].

[60] D.Y. Chen, Q.Q Jiang and X.T. Zu, Class. Quant. Grav. 25 (2008)
205022; Phys. Lett. B665 (2008) 106.

[61] Q.Q. Jiang, Phys. Rev. D78 (2008) 044009; Phys. Lett. B666
(2008) 517.

[62] K. Lin and S.Z. Yang, Phys. Rev. D79 (2009) 064035; Phys. Lett.
B674 (2009)127.

[63] H.L. Li, W.Y. Qi and R. Lin, Phys. Lett. B677 (2009) 332.

\end{document}